\documentclass[sigconf]{acmart}
\usepackage{algorithmic}
\usepackage{graphicx}
\usepackage{textcomp}
\usepackage{graphicx}
\usepackage{caption}
\usepackage{subcaption}
\usepackage{bbm}
\usepackage{color}
\usepackage{hyperref}
\usepackage{multirow}
\usepackage{enumitem}
\usepackage{algorithm} 
\usepackage{algorithmic}
\usepackage{enumerate}
\usepackage{amsmath}
\usepackage{booktabs}
\usepackage{multirow}
\usepackage{stmaryrd}
\usepackage{mathrsfs} 
\usepackage{color}
\usepackage{adjustbox}
\usepackage{balance}

\newcommand{\eg}{\textit{e.g.}}

\newcommand{\ie}{\emph{i.e.}}

\AtBeginDocument{%
  }
\copyrightyear{2026}
\acmYear{2026}
\setcopyright{cc}
\setcctype{by}
\acmConference[SIGIR '26]{Proceedings of the 49th International ACM SIGIR Conference on Research and Development in Information Retrieval}{July 20--24, 2026}{Melbourne, VIC, Australia}
\acmBooktitle{Proceedings of the 49th International ACM SIGIR Conference on Research and Development in Information Retrieval (SIGIR '26), July 20--24, 2026, Melbourne, VIC, Australia}
\acmDOI{10.1145/3805712.3809681}
\acmISBN{979-8-4007-2599-9/2026/07}

\begin{document}

\title{Modular Representation Compression: Adapting LLMs for Efficient and Effective Recommendations}




\author{Yunjia Xi}
\email{xiyunjia@sjtu.edu.cn}
\affiliation{%
  \institution{Shanghai Jiao Tong University}
  \city{Shanghai}
  \country{China}
}

\author{Menghui Zhu}
\email{zhumenghui1@huawei.com}
\affiliation{%
  \institution{Huawei Noah's Ark Lab}
  \city{Shanghai}
  \country{China}
}

\author{Jianghao Lin}
\authornote{Corresponding authors.}
\email{linjianghao@sjtu.edu.cn}
\affiliation{%
  \institution{Antai College of Economics and Management, Shanghai Jiao Tong University}
  \city{Shanghai}
  \country{China}
}

\author{Bo Chen}
\email{chenbo116@huawei.com}
\affiliation{%
  \institution{Huawei Noah's Ark Lab}
  \city{Shanghai}
  \country{China}
}

\author{Ruiming Tang}
\email{tangruiming@huawei.com}
\affiliation{%
  \institution{Huawei Noah's Ark Lab}
  \city{Shenzhen}
  \country{China}
}

\author{Yong Yu}
\email{yyu@sjtu.edu.cn}
\affiliation{%
  \institution{Shanghai Jiao Tong University}
  \city{Shanghai}
  \country{China}
}

\author{Weinan Zhang}
\email{wnzhang@sjtu.edu.cn}
\affiliation{%
  \institution{Shanghai Jiao Tong University}
  \city{Shanghai}
  \country{China}
}

\renewcommand{\shortauthors}{Yunjia Xi et al.}

\begin{abstract}
  Recently, large language models (LLMs) have advanced recommendation systems (RSs), and recent works have begun to explore how to integrate LLMs into industrial RSs. While most approaches deploy LLMs offline to generate and pre-cache augmented representations for RSs, high-dimensional representations from LLMs introduce substantial storage and computational costs. Thus, it is crucial to compress LLM representations effectively. However, we identify a counterintuitive phenomenon during representation compression: \textbf{Mid-layer Representation Advantage (MRA)}, where representations from middle layers of LLMs outperform those from final layers in recommendation tasks. This degraded final layer renders existing compression methods, which typically compress on the final layer, suboptimal.
We interpret this based on modularity theory that LLMs develop spontaneous \textbf{internal functional modularity} and force the final layer to specialize in the proxy training task. Thus, we propose \underline{M}odul\underline{a}r \underline{R}epresentation \underline{C}ompression (\textit{MARC}) to explicitly control the modularity of LLMs. First, \textit{Modular Adjustment} explicitly introduces compression and task adaptation modules, enabling the LLM to operate strictly as a representation-learning module. Next, to ground each module to its specific task, \textit{Modular Task Decoupling} uses information constraints and different network structures to decouple tasks. Extensive experiments validate that MARC addresses MRA and produces efficient representations. Notably, MARC achieved a 2.82\% eCPM lift in an online A/B test within a large-scale commercial search advertising scenario. 
\end{abstract}








\maketitle

\section{Introduction}
The rapid advancement of large language models (LLMs) has profoundly transformed numerous domains, including recommendation systems (RSs)~\cite{zhu2024lifelong,yu2023self,lin2025can}. Recent studies have increasingly explored the integration of LLMs into RSs, achieving performance that surpasses traditional recommendation approaches~\cite{fan2023recommender,wu2023survey,liu2023pre,lin2024rella}. However, their deployment in industrial RSs remains at a nascent stage. This is primarily because commercial RSs typically handle millions of users and items while adhering to \textit{stringent response latency requirements} -- often within 100ms~\cite{xi2023device,lin2025large,liu2024behavior}. The substantial computational overhead and significant inference latency of LLMs render them impractical for online serving in such scenarios. Consequently, most deployment strategies \textbf{pre-cache LLM representations for traditional RSs}, avoiding their online inference~\cite{xi2025efficiency,luo2024kellmrec,xi2023towards,ren2024representation}. A prevalent pipeline entails leveraging LLMs offline to generate or encode textual knowledge into knowledge representations, which are subsequently utilized by downstream recommendation models.

\begin{figure}[]
\vspace{5pt}
    \centering
    \begin{subfigure}{0.23\textwidth}
        \centering
        \includegraphics[width=\linewidth]{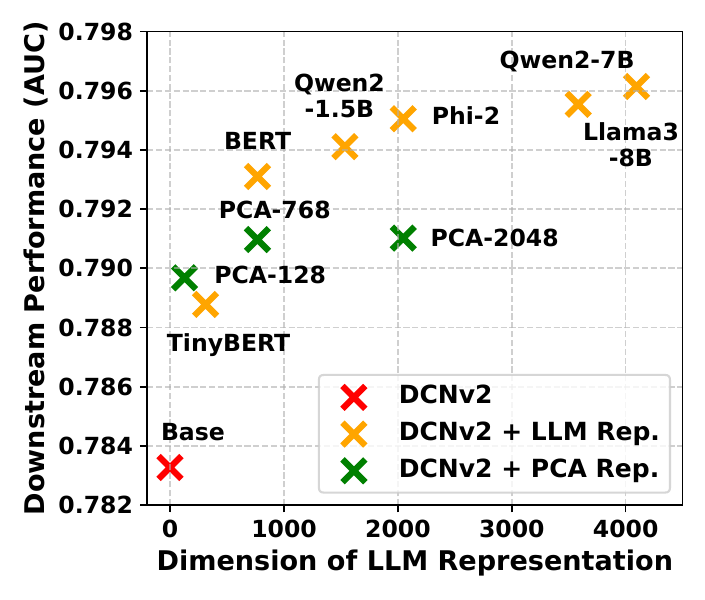}
        \vspace{-12pt}
        \caption{Performance}
        \label{fig:intro_a}
    \end{subfigure}
    \begin{subfigure}{0.23\textwidth}
        \centering
        \includegraphics[width=\linewidth]{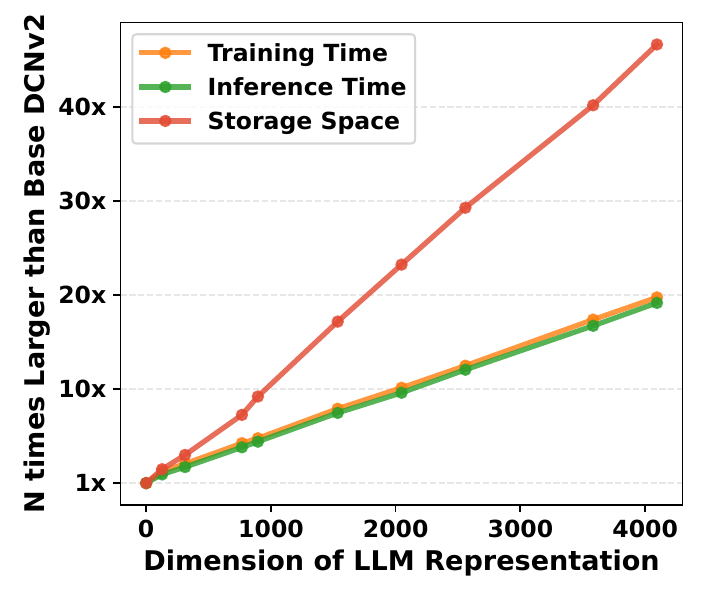}
        \vspace{-12pt}
        \caption{Costs}
        \label{fig:intro_b}
    \end{subfigure}
    \vspace{-5pt}
    \caption{The impact of various LLM representations on downstream performance and costs in MovieLens-1M.}
    \label{fig:intro}
    \vspace{-10pt}
\end{figure}

The LLM representations can inject richer information into RSs and significantly enhance their performance~\cite{luo2024kellmrec,xi2023towards,ren2024representation}, but there is also a \textbf{trade-off between effectiveness and efficiency}. In Figure~\ref{fig:intro_a}, larger LLMs bring greater enhancements to the downstream Click-Through Rate (CTR) model DCNv2~\cite{wang2021dcnv2} (\eg, BERT~\cite{kenton2019bert} < Phi-2~\cite{javaheripi2023phi} < Llama3-8B~\cite{dubey2024llama}). However, larger LLMs produce higher-dimensional representations, substantially increasing storage, training, and inference overhead (Figure~\ref{fig:intro_b}), as detailed in Section~\ref{sec:llm_size}. In industrial RSs, where user and item volumes far exceed public datasets, these costs are more prohibitive. While simple dimensionality reduction methods like PCA~\cite{mackiewicz1993pca} can reduce costs, they suffer from severe performance degradation. 
In Figure~\ref{fig:intro_a}, when PCA reduces Llama3-8B's representation to BERT's dimensionality (PCA-768), it performs worse than BERT.
This suggests that \textit{training-free compression} methods, such as PCA, struggle to retain information useful for downstream tasks, making fine-tuning LLMs essential for efficient representation compression.

Existing fine-tuning-based compression methods, including nested-based~\cite{kusupati2022matryoshka,wang20242d} and projection-based~\cite{wan2024larr} approaches, have significantly outperformed training-free methods. These methods typically compress on the final-layer output of LLMs with training objectives like contrastive loss. However, in our experiments~\ref{sec:mra}, we identify a pervasive and critical issue: \textbf{Mid-layer Representation Advantage (MRA)}, where \textit{representations from middle layers outperform those from the final layer in recommendation tasks}. This observation reveals a fundamental limitation: existing training paradigms for representation compression degrade the quality of final-layer representations. Consequently, methods that focus on compressing this suboptimal final layer are bound to limit their potential effectiveness.



We interpret this phenomenon through the lens of modularity theory that LLMs develop spontaneous \textbf{internal functional modularity} during training. Specifically, the early-to-middle layers function as a \textit{Representation Learning Module}, extracting rich, generalized textual features. However, a critical functional shift occurs in the final layers. Under current training paradigms, the strong supervision signals from the training objective (\eg, contrastive loss) force the final layers to evolve into a specialized \textit{Task Adaptation Module}. These final layers form an unintended \textbf{information bottleneck}, filtering out diverse semantics deemed redundant for the training objective but crucial for recommendation. This explains the MRA: while middle layers retain generalized information, the final layer sacrifices it for task adaptation. This emergent modularity is difficult to predict a priori, making it challenging to directly compress the optimal middle-layer representations.

Based on this insight, we propose a \underline{M}odul\underline{a}r \underline{R}epresentation \underline{C}ompression (dubbed \textit{MARC}) framework to transform this implicit, uncontrolled modularity into an explicit, controllable architectural design. Our core intuition is to preserve the high-quality representations in the final layer by structurally decoupling representation learning from task adaptation. \textit{First}, we perform \textbf{Modular Adjustment} by explicitly introducing external modules responsible for different tasks. We introduce a lightweight \textit{Compression Network} to act as the dedicated compression module and a \textit{User-Item Matching Network} to offload the task adaptation burden. This explicit separation prevents the LLM's final layers from collapsing into task-specific heads, allowing the backbone to focus purely on learning high-quality, generalized representations. \textit{Next}, to ensure that each module performs its specific task, we introduce \textbf{Modular Task Decoupling}, including information constraint, Hilbert-Schmidt Independence Criterion (HSIC), and different network structures for different tasks. HSIC constraint guides compression by maximizing the mutual information between original and compressed representations, effectively separating representation and compression tasks. The compression network processes individual representations, while the matching network inputs both user and item representations, capturing their explicit and implicit interactions. These distinct structures separate compression and target tasks. This framework ensures that information filtering occurs strictly within the external modules, leaving the LLM's representation capability intact. Our contributions can be summarized as follows:
\begin{itemize}
    \item We identify the tradeoff between efficiency and effectiveness in LLM-based representations for recommendation. To the best of our knowledge, this is the \textit{first work that focuses on representation compression in LLM-based recommendation}. 
    \item We reveal the \textbf{mid-layer representation advantage} issue in common LLM representation compression methods, and we analyze and explain it with modular theory for the first time.
    \item We propose MARC to address this issue, with Modular Adjustment to add compression and user-item matching network and Modular Task Decoupling to ensure each module performs its specific task, enabling LLMs to focus on representation learning. 
    \item Experiments validate that MARC mitigates mid-layer representation advantage and improves eCPM by 2.82\% during an \textbf{online A/B test} in a large commercial search advertising scenario.
\end{itemize}

\section{Analyses on LLM Representations for RSs}
\subsection{Background: LLM Representations for RSs}
LLMs' massive parameters and autoregressive nature bring high inference latency, making it challenging to meet the online low-latency requirements of RSs. As a result, currently deployable LLM-based recommendation solutions typically leverage LLMs offline to generate representations for users and items, which can then be utilized in traditional recommendation tasks~\cite{xi2023towards,ren2024representation,zhang2024notellm,hu2024enhancing,luo2024kellmrec}. We refer to this approach as ``LLM Representations for RSs''. 

Given the item set $\mathcal{I}$ and the user set $\mathcal{U}$, each item $i\in\mathcal{I}$ and user $u\in\mathcal{U}$ possesses specific attributes and descriptions, which can be converted into text $T_i$ and $T_u$ through predefined templates.  
In some  work~\cite{luo2024kellmrec}, the text can incorporate LLM-generated knowledge to infer user preferences. 
Subsequently, $T_i$ and $T_u$ are transformed into representations $r_i$ and $r_u$, with dimension $d_o$, through LLMs:
\begin{equation}\label{eq:encoding}
    r_i=\text{Pooling}(\text{LLM}(T_i)), \,\,\,r_u=\text{Pooling}(\text{LLM}(T_u)),
\end{equation}
where LLMs may be pre-trained or fine-tuned models, and pooling methods like mean pooling or EOS token pooling are used to obtain the final representations. Then, the user and item representations can be utilized as static features in any downstream models, \eg, click-through rate (CTR) prediction models.

\subsection{Tradeoff between Performance and Costs}\label{sec:llm_size}
\begin{figure}[htbp]
    \centering
    \vspace{-5pt}
    \begin{subfigure}{0.23\textwidth}
        \centering
        \includegraphics[width=\linewidth]{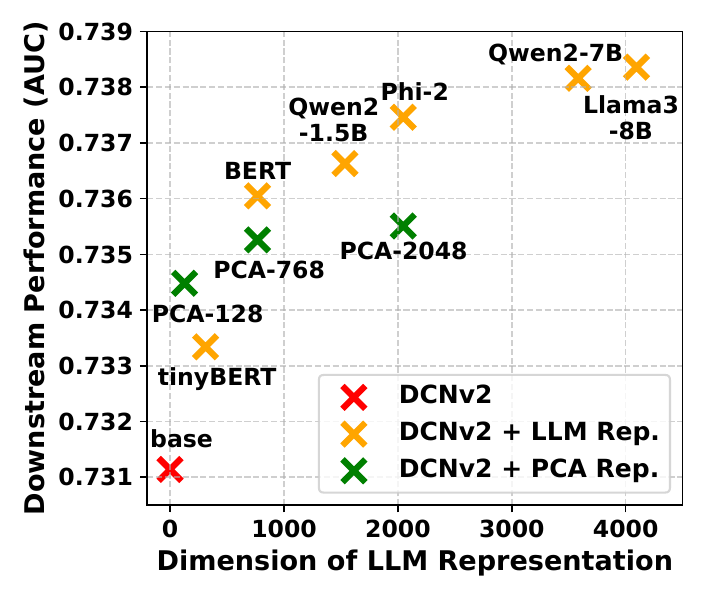}
        \caption{Performance}
        \label{fig:intro_a_yelp}
    \end{subfigure}
    \begin{subfigure}{0.23\textwidth}
        \centering
        \includegraphics[width=\linewidth]{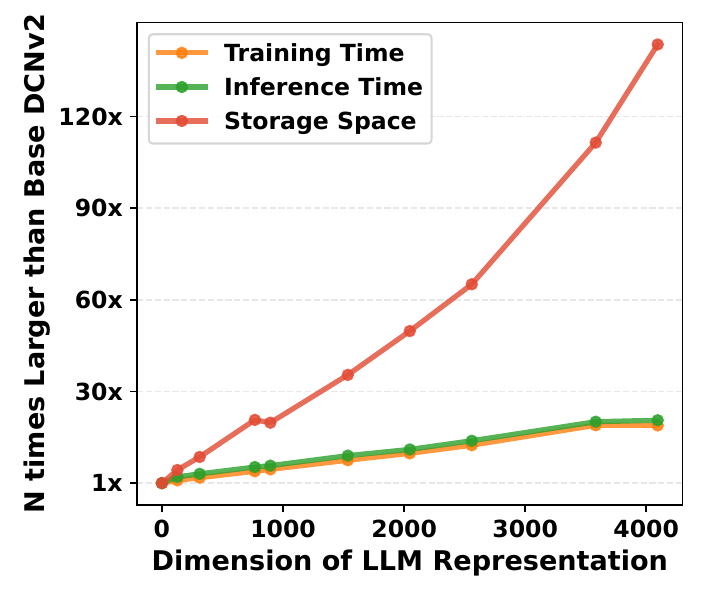}
        \caption{Costs}
        \label{fig:intro_b_yelp}
    \end{subfigure}
    \vspace{-5pt}
    \caption{The impact of different sizes of LLMs on downstream DCNv2 performance and costs in Yelp.}
    \label{fig:impact_yelp}
    \vspace{-5pt}
\end{figure}

We first empirically investigate the impact of LLM scaling on recommendation performance on MovieLens-1M and Yelp datasets following~\cite{xi2023towards}. First, item-related features (e.g., titles) and user-related features (e.g., the most recent history) are converted to text using templates. Next, these texts are encoded into representations by LLMs, and then, the representations are incorporated into the downstream CTR model DCNv2~\cite{wang2021dcnv2} (\textbf{Base}). The LLMs used in this study are pre-trained models without fine-tuning, including a range of models from small to large: \textbf{TinyBERT}~\cite{jiao2019tinybert}, \textbf{BERT}~\cite{kenton2019bert}, \textbf{Phi-2}~\cite{javaheripi2023phi}, \textbf{Qwen2-1.5B/7B}~\cite{hui2024qwen2}, and \textbf{Llama3-8B}~\cite{dubey2024llama}. Additionally, for Llama3-8B, its 4096-dimensional representations are reduced to 128, 768, and 2048 dimensions with PCA, producing \textbf{PCA-128}, \textbf{PCA-768}, and \textbf{PCA-2048}, respectively. The results on two datasets are presented in Figure~\ref{fig:intro} and Figure~\ref{fig:impact_yelp}. 


As illustrated in Figure~\ref{fig:intro_a_yelp}, there is a clear positive correlation between model size and downstream performance. \textit{Larger LLMs consistently yield higher performance, but this comes at a steep cost}. In Figure~\ref{fig:intro_b_yelp}, the storage, training, and inference overheads (which are expressed as multiples of base DCNv2) grow linearly with the representation dimension. For instance, deploying 4096-dimensional representations from Llama3-8B incurs a 140$\times$ increase in storage and 20$\times$ increase in training time compared to base settings. While simple dimension-reduction methods like PCA can reduce costs, they suffer from significant performance degradation. Thus, we need advanced representation compression techniques tailored to recommendations.


\begin{figure*}[h]
    \centering
    \vspace{-0pt}
    \includegraphics[width=1\textwidth]{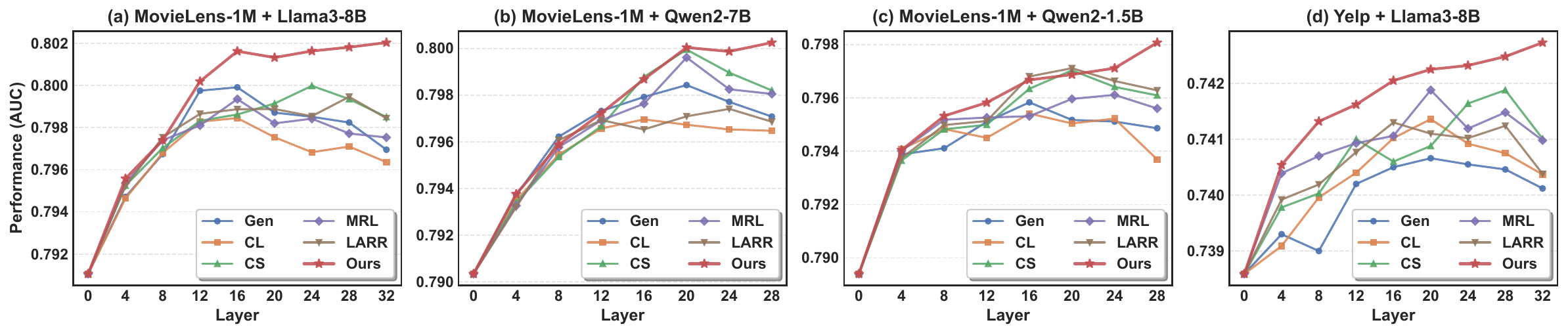}
    \vspace{-5pt}
    \caption{Mid-layer representation advantage across different datasets and LLMs.}
    \vspace{-5pt}
    \label{fig:mra}
\end{figure*}

\subsection{Mid-layer Representation Advantage (MRA) During Compression}\label{sec:mra}
To achieve effective compression, recent works employ \textbf{fine-tuning-based compression methods} (e.g., MRL~\cite{kusupati2022matryoshka}, LARR~\cite{wan2024larr}), which all require fine-tuning on specific \textit{proxy tasks}, such as contrastive loss, to achieve effective compression. Thus, we fine-tune Llama3-8B, Qwen2-7B, and Qwen2-1.5B on the MovieLens-1M and Yelp dataset and explore the layer-wise performance of various proxy target tasks, including both compression and non-compression approaches: (1) \textbf{Gen} employs next-token prediction to predict the next positive item based on a user’s history; (2) \textbf{CL} uses in-batch negative samples for contrastive learning; (3) \textbf{CS} uses cosine similarity loss between user and item representations; (4) \textbf{MRL}~\cite{kusupati2022matryoshka}, nested-based compression with cross-entropy (CTR) loss; (5) \textbf{LARR}~\cite{wan2024larr}, projection-based compression with contrastive learning; (6) \textbf{Ours}, our proposed MARC. All the above models utilize the same click data for training. Then, we encoded user and item texts into representations with the fine-tuned LLMs. Lastly, representations from different layers of LLMs are applied to the downstream DCNv2.


With the exception of our proposed MARC, \textbf{the representations that yield the best downstream performance consistently originate from the middle-to-later layers rather than the final layer}. We term this phenomenon the Mid-layer Representation Advantage. Specifically, while performance steadily increases in the early layers, it peaks in a middle layer and then drops as it approaches the final layer. Crucially, MRA persists even when baselines are fine-tuned with the exact same recommendation-aligned loss (CTR loss for MRL), confirming \textit{MRA is a structural characteristic of LLMs in RS rather than an objective misalignment issue}. 

This highlights a fundamental deficiency in current compression methods: their training paradigms degrade the quality of final-layer representations. Moreover, layer-wise selection in Figure~\ref{fig:mra} is impractical in real-world scenarios as the optimal layer varies across models and tasks. Consequently, existing compression methods, which operate on the final layer, are inherently suboptimal. In contrast, our MARC is designed to prevent this degeneration, and its final representation significantly outperform its middle layers and even the best-performing intermediate layers of all baselines.

\vspace{-5pt}
\subsection{Understanding MRA: A Modular Perspective}\label{sec:Modularity}
Why does the final layer, theoretically the most ``processed'' one, fail to yield the optimal performance? We interpret it within the framework of \textbf{Functional Modularity} and \textbf{Layer-wise Specialization}~\cite{xiao2024configurable,zhang2023emergent}. Recent studies on the interpretability of Transformers have observed \textit{emergent modularity}~\cite{xiao2024configurable,zhang2023emergent}, where pre-trained LLMs spontaneously differentiate into functional regions specialized in different tasks during training. 

Building on these findings, we argue that during the previous fine-tuning process for representation compression, the LLM spontaneously develops two internal functional modules:
\begin{itemize}
    \item \textbf{Representation Learning Module (Early-to-Middle Layers)}: The layers from the bottom up to the peak of MRA primarily focus on understanding and extracting rich, generalized features, capturing diverse information potentially useful for various tasks.
    \item \textbf{Task Adaptation Module (Final Layers)}: These layers evolve to adapt these features to the training objective, \eg, contrastive loss, due to their proximity to the supervision signal. 
\end{itemize}
Crucially, in existing training paradigms, \textbf{the boundary between these two modules is spontaneous and uncontrollable}. Even when projection heads~\cite{wan2024larr} are employed, they fail to shield the backbone. The strong supervision signals from the training loss inevitably propagate back to the LLM's final layers, turning them into Task Adaptation Module. 
This aligns with the Information Bottleneck principle~\cite{tishby2015deep} that deep networks compress input data into a minimal sufficient statistic for the target. Thus, the final layers filter out information redundant for the proxy objective but vital for recommendations.

\begin{figure}[htbp]
    \centering
    \vspace{0pt}
    \includegraphics[width=0.43\textwidth]{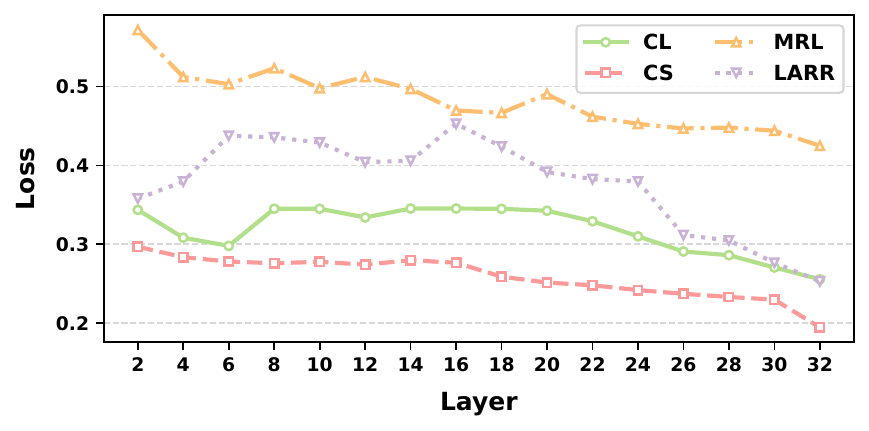}
    \vspace{-5pt}
    \caption{Layer-wise performance on the proxy target task.}
    \vspace{-5pt}
    \label{fig:loss}
\end{figure}

\paragraph{Evidence from Target Task Analysis.} To empirically validate this functional partitioning, we analyze the layer-wise performance on the fine-tuning proxy task itself. We use the representations extracted from each layer to calculate the final optimization loss, which serves as an indicator of their ability to handle the proxy target task. As illustrated in Figure~\ref{fig:loss}, the optimization loss on the proxy tasks (CL, CS) decreases consistently, reaching its minimum at the final layer. This provides compelling evidence that the \textit{later layers are highly specialized for the proxy training objective}, functioning as the Task Adaptation Module. However, this comes at a cost: the model progressively discards information deemed redundant for the proxy task but critical for recommendations.

\paragraph{Difficulty in Aligning Proxy Targets with Recommendation.} Real-world recommendation scenarios are \textit{multifaceted}, encompassing diverse tasks such as retrieval, ranking, and re-ranking, which thrive on rich, multi-dimensional semantic information. In contrast, fine-tuning tasks are typically driven by \textit{single-objective proxies}. Thus, useful representations for recommendations are difficult to perfectly align with any single proxy target. Crucially, our experiments (Figure~\ref{fig:mra}) show that MRA persists even when domain-aligned objectives like CTR loss are used in MRL. Consequently, the spontaneous modularity driven by single-objective supervision force the final layer to specialize in proxy target and lose rich information vital for RSs. Compressing an already degraded final layer inevitably leads to suboptimal performance, necessitating a new framework that can explicitly control the modularity.

\vspace{-4pt}
\section{Modular Representation Compression}
\subsection{Overview}
\begin{figure}[htbp]
    \centering
    \vspace{-5pt}
    \includegraphics[width=0.52\textwidth, trim=18 0 0 0, clip]{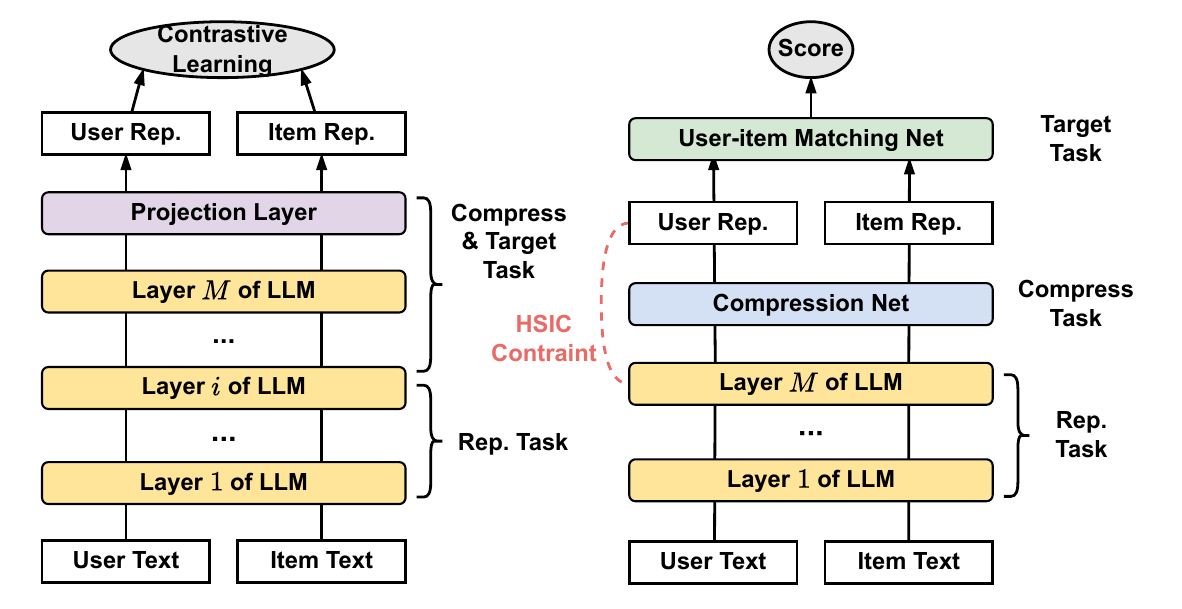}
    \vspace{-5pt}
    \caption{Comparison between existing projection-based compression methods (left) and MARC (right). Note that User and Item Representation (Rep.) are used for downstream tasks.}
    \vspace{-5pt}
    \label{fig:fw}
\end{figure}

Based on the above findings and analysis, we propose MARC to transform this implicit, uncontrolled modularity into an explicit, controllable architectural design. Existing nested-based and projection-based compression methods lack explicit constraints on functional regions, so under the supervision of training tasks, the later layers of LLMs are still involved in other functions, \eg, task adaptation, as shown in the left panel of Figure~\ref{fig:fw}. This leads to MRA in downstream tasks, as discussed in Section~\ref{sec:Modularity}. MARC, in the right panel of Figure~\ref{fig:fw}, introduces \textbf{explicit modularity}, specifically two external modules for compression and task adaptation and \textit{enforcing clear modular boundaries}. This ensures that LLM's final layers retain their robust representation learning capabilities.

Specifically, we make \textbf{Modular Adjustment}, retaining the LLM strictly as a representation learning module, while introducing two external lightweight modules: (1) a compression network to condense high-dimensional semantics, and (2) a user-item matching network to serve as the dedicated task adaptation module. This design effectively prevents the LLM's final layers from collapsing into task-specific heads, thereby preserving the rich, generalized information (typical of middle layers) throughout the entire network depth. To ensure functional decoupling, we devise \textbf{Modular Task Decoupling}. The information constraint HSIC is leveraged to maximize the mutual information between original and compressed representations, improving the information density of the compressed representation. The compression network processes individual representations, while the matching network inputs both user and item representations and models their interactions. These distinct structures separate compression and target tasks while preserving essential interactive information for recommendations. The model pipeline includes three modules: a representation module with LLMs, a compression module with a compression network, and a task adaptation module via a user-item matching network, and \textbf{these modules are trained in an end-to-end manner}.

\subsection{Representation Module}
This stage is handled by LLMs, fully leveraging their powerful representational capabilities. Similar to previous works, our model takes as input the text related to the user $u$ and the item $i$. Each item $i$ has its attributes (\eg, category) and description, which are transformed into text $T_i$ through templates. Each user $u$ also has specific attributes (\eg, age, gender) and historical behaviors $H_u$, which are also converted into text $T_u$ with templates. These texts are then encoded by LLMs. After pooling the representation from LLMs' last layer, we obtain the user and item representations $r_u\in\mathbb{R}^{d_o}$ and $r_i\in\mathbb{R}^{d_o}$ of dimension $d_o$, following Eq.~\eqref{eq:encoding}. To distinguish representations generated in this stage from those in the compression stage, we refer to $r_u$ and $r_i$ as the \textbf{original representations}. 

Crucially, we deviate from standard fine-tuning paradigms, which typically impose task-specific losses directly on the LLM's output representations $r_u$ and $r_i$. Such direct supervision forces the LLM's final layers to over-fit the specific target loss, turning the backbone into a task-specific predictor. In contrast, \textbf{MARC does not apply any direct supervision to the raw LLM representations $r_u$ and $r_i$ at this stage}. Instead, the task-specific loss is imposed strictly on the subsequent lightweight modules (\ie, the Compression and Matching Networks). Consequently, the LLM backbone is optimized indirectly via backpropagation through these external modules. This architectural design ensures that the task adaptation pressure is primarily absorbed by the external modules, allowing the LLM backbone to maintain generalized, high-quality representation without collapsing into a narrow, task-specific subspace.

\subsection{Compression Module}
For compression, we introduce a dedicated compression network, $g(\cdot)$, which takes user and item representations, $r_u\in\mathbb{R}^{d_o}$ and $r_i\in\mathbb{R}^{d_o}$, and generates the \textbf{compressed representations} $c_u\in\mathbb{R}^{d_c}$ and $c_i\in\mathbb{R}^{d_c}$ of dimension $d_c (d_c\ll d_o)$, as follows:
\begin{equation}
    c_i=g(r_i),\,\, c_u=g(r_u).
\end{equation}
For simplification, we take the user's original representation $r_u$ and compressed representation $c_u$ as examples to introduce the information constraint. The same process can be applied to item representations. During compression, we aim to preserve as much of the information from the original representations as possible in the compressed ones, which means maximizing the mutual information (MI) between the original and compressed representations:
\begin{equation}
    \max_\theta MI(r_u, c_u),
\end{equation}
where $\theta$ represents the model parameters, and $MI(\cdot)$ denotes the mutual information between the two representations. Common mutual information optimization methods, such as MSE, InfoNCE~\cite{oord2018representation}, and DIM~\cite{hjelm2018learning}, require the two representations to have matching dimensions. However, $r_u$ and $c_u$ reside in spaces of vastly different conceptual levels and dimensions, and mapping them to the same dimension with a simple projection layer could distort the information. To ensure the Compression Network retains the maximum amount of information from the backbone, we employ the sampling form of mutual information, \textbf{Hilbert-Schmidt Independence Criterion} (HSIC)~\cite{gretton2005measuring,ma2020hsic}, as the constraint. It utilizes a kernel function to map vectors of different dimensions into a higher-dimensional Reproducing Kernel Hilbert Space (RKHS). This enables us to handle the \textbf{dimension mismatch} and maximize \textbf{non-linear dependencies} between the original and compressed views, effectively transferring rich information to the lower-dimensional space without requiring dimension-matching projections. 
Mathematically, given a data batch $D=\{(x_1,y_1), \ldots, (x_n, y_n)\}$ ($n$ denotes batch size) from two random variables $X$ and $Y$, the HSIC is defined as follows:
\begin{equation}
    \mathcal{L}_{HSIC}=HSIC(X,Y)=\frac{1}{(n-1)^2}Tr(K_XJK_YJ),
\end{equation}
where $K_X\in\mathbb{R}^{n\times n}$and $K_Y\in\mathbb{R}^{n\times n}$ are the kernel matrices generated by applying kernel functions to the samples from $X$ and $Y$. Here,each entry of $K_X$ is defined as:
\begin{equation}
    K_X(i,j)=k_X(x_i,x_j),\,\, K_Y(i,j)=k_Y(y_i,y_j),
\end{equation}
where $k_X(\cdot)$ and $k_Y(\cdot)$ are the kernel functions for $X$ and $Y$, respectively. Typically, a Gaussian kernel is employed for them, \ie,
\begin{equation}
    k_X(x_i,x_j)=e^{-\frac{||x_i-x_j||^2}{2\sigma^2}},
\end{equation}
where $\sigma$ is a hyperparameter. The matrix $J=I_n-\frac{1}{n}I_nI_n^T$ is the centering matrix, which removes biases in the data by ensuring that the kernel matrices are centered. The operator $Tr(\cdot)$ denotes the trace of a matrix, summarizing the correlation of variables in the higher-dimensional space. It is worth noting that the computational complexity of HSIC is $O(n^2)$ where $n$ is the batch size. This is negligible compared to computation of the LLM backbone.

This constraint achieves two key objectives: \textit{preserving relevant information} during compression and \textit{effectively decoupling tasks}. First, it maximizes the mutual information between the compressed and original representations, ensuring that the compressed representations retain as much information as possible and improving information density. Second, it separates the compression and representation tasks, reinforcing the distinct functions of the compression network and LLMs' representation learning. This enhances the compression functionality of the compression network while maintaining the representational power of LLMs. 

\subsection{Task Adaptation Module}
We introduce the User-Item Matching Network as the dedicated \textit{Task Adaptation Module} to serve as the designated sink for supervision signals, effectively absorbing the optimization pressure from the training objective. Unlike simple projection heads~\cite{wan2024larr}, this network incorporates rich explicit and implicit interaction mechanisms that are more aligned with downstream recommendation tasks, allowing it to naturally assume the adaptation burden and clearly distinguishing them from the compression phase. 


Specifically, we introduce both \textbf{explicit and implicit interaction}. The input representations $(c_u, c_i)$ first undergo explicit manual feature interactions ($|c_i-c_u|$ and $c_i\odot c_u$) before passing through the interaction network $f(\cdot)$ for implicit interaction. This network predicts whether user $u$ likes item $i$, denoted as $\hat{y}$, 
\begin{equation}
    \hat{y} = f(c_i,c_u,|c_i-c_u|,c_i\odot c_u),
    \label{eq:matching_net}
\end{equation}
where $\odot$ denotes the element-wise product. This distinction in input and network structure helps better separate the matching and compression stages. These explicit and implicit interactions are better aligned with recommendation tasks, enabling representations to retain more information relevant to recommendations and thereby improving representation quality. This structure ensures that the information bottleneck, filtering redundant information, and focusing on the target occur strictly within this lightweight external module, preventing the LLM representations from collapsing into task-specific logits. The matching task is designed with a cross-entropy loss function:
\begin{equation}
    \mathcal{L}_{match}=-\sum_{(i,u,y)\in D'}y\log\hat{y}+(1-y)\log (1-\hat{y}),
\end{equation}
where $D'$ denotes the training data set and $y$ represents the ground truth label (typically whether the item $i$ is clicked or highly rated by user $u$). The representation, compression, and task adaptation modules of MARC are optimized in an \textbf{end-to-end manner} with the final loss function:
\begin{equation}
    \mathcal{L}=\mathcal{L}_{match} + \alpha\mathcal{L}_{HSIC}.
    \label{eq:loss}
\end{equation}

Upon completing the end-to-end training, we freeze the model to produce stable, low-dimensional representations, which are leveraged as static features for downstream recommendation models.

\section{Experiment}

To gain more insights into our proposed MARC, we aim to address the following research questions (RQs) in this section. 
\begin{itemize}
    \item \textbf{RQ1:} How does our model, MARC, perform on various downstream recommendation tasks?
    \item \textbf{RQ2:} What is the online performance of MARC?
    \item \textbf{RQ3:} What role does each module of MARC play?
    \item \textbf{RQ4:} Is MARC compatible with downstream models and LLMs?
    \item \textbf{RQ5:} How does our model perform with different sizes of compressed representations?

\end{itemize}
\subsection{Setup}

\begin{table*}[htbp]

    \caption{The overall performance on CTR prediction task. The best result is given in bold, while the second-best compression baseline is underlined. MARC-C and MARC-O represent the \underline{c}ompressed and the \underline{o}riginal representations of our MARC. MARC-O is the upper bound of MARC-C. Rel. Impr denotes the relative AUC improvement of MARC-C against other models. The symbol * indicates statistically significant improvement over the best compression baselines (t-test with $p < 0.05$).}
    \centering
    \scalebox{0.96}{
    \setlength{\tabcolsep}{1.5mm}
{\begin{tabular}{c|c|c|ccc|ccc|ccc}
\toprule
\multirow{2}{*}{\begin{tabular}[c]{@{}c@{}}Pre-trained\\ Language Models\end{tabular}} & \multirow{2}{*}{\begin{tabular}[c]{@{}c@{}}Compression\\ Method\end{tabular}} & \multirow{2}{*}{Dimension} & \multicolumn{3}{c|}{ML-1M} & \multicolumn{3}{c|}{ML-25M} & \multicolumn{3}{c}{Yelp} \\
\cmidrule{4-12}
 &  &  & AUC & Logloss & Rel.Impr & AUC & Logloss & Rel.Impr & AUC & Logloss & Rel.Impr \\
 \midrule
base (DCNv2) & / & / & 0.7833 & 0.5512 & 2.28\% & 0.7945 & 0.5469 & 3.63\% & 0.7311 & 0.5203 & 1.58\% \\
TinyBERT & / & 312 & 0.7888 & 0.5459 & 1.57\% & 0.8079 & 0.5321 & 1.91\% & 0.7333 & 0.5173 & 1.27\% \\
BERT & / & 768 & 0.7931 & 0.5418 & 1.01\% & 0.8119 & 0.5276 & 1.41\% & 0.7361 & 0.5159 & 0.90\% \\
Qwen2-1.5B & / & 1536 & 0.7941 & 0.5410 & 0.88\% & 0.8141 & 0.5253 & 1.14\% & 0.7366 & 0.5153 & 0.82\% \\
Phi-2 & / & 2560 & 0.7954 & 0.5390 & 0.73\% & 0.8201 & 0.5178 & 0.40\% & 0.7378 & 0.5145 & 0.67\% \\
Qwen2-7B & / & 3584 & 0.7955 & 0.5402 & 0.71\% & 0.8213 & 0.5153 & 0.25\% & 0.7382 & 0.5141 & 0.61\% \\
\midrule
\multirow{4}{*}{\begin{tabular}[c]{@{}c@{}}Llama3-8B\\ (frozen)\end{tabular}} & PCA & 128 & 0.7897 & 0.5449 & 1.45\% & 0.8138 & 0.5222 & 1.18\% & 0.7345 & 0.5175 & 1.12\% \\
 & AE & 128 & 0.7893 & 0.5478 & 1.50\% & 0.8145 & 0.5214 & 1.08\% & 0.7329 & 0.5172 & 1.34\% \\
 & VAE & 128 & 0.7840 & 0.5504 & 2.18\% & 0.8094 & 0.5298 & 1.72\% & 0.7323 & 0.5177 & 1.42\% \\
 & / & 4096 & 0.7961 & 0.5397 & 0.63\% & 0.8215 & 0.5157 & 0.23\% & 0.7384 & 0.5137 & 0.59\% \\
 \midrule
\multirow{6}{*}{\begin{tabular}[c]{@{}c@{}}Llama3-8B\\ (fine-tuned )\end{tabular}} & LARR & 128 & 0.7911 & 0.5444 & 1.27\% & 0.8179 & 0.5219 & 0.67\% & 0.7354 & 0.5170 & 0.99\% \\
 & BAHE & 128 & 0.7936 & 0.5408 & 0.95\% & 0.8182 & 0.5191 & 0.63\% & 0.7360 & 0.5150 & 0.91\% \\
 & ESE & 128 & 0.7940 & 0.5406 & 0.90\% & 0.8183 & 0.5193 & 0.61\% & 0.7364 & \underline{0.5147} & 0.85\% \\
 & MRL & 128 & \underline{0.7943} & \underline{0.5404} & 0.86\% & \underline{0.8186} & \underline{0.5188} & 0.58\% & \underline{0.7372} & 0.5166 & 0.74\% \\
 & MARC-C & 128 & \textbf{0.8011*} & \textbf{0.5317*} & - & \textbf{0.8233*} & \textbf{0.5127*} & - & \textbf{0.7427*} & \textbf{0.5109*} & - \\
 \cmidrule{2-12}
 & \begin{tabular}[c]{@{}c@{}}MARC-O\\ (\textit{oracle})\end{tabular} & 4096 & 0.8020* & 0.5321* & -0.11\% & 0.8254* & 0.5100* & -0.25\% & 0.7428* & 0.5105* & -0.01\%\\
 \bottomrule
\end{tabular}}
}
\label{tab:overall}
\end{table*}

\subsubsection{Datasets and Pipelines} The experiments are conducted on three public datasets, MovieLens-1M\footnote{\url{https://grouplens.org/datasets/movielens/1m/}}, Yelp\footnote{\url{https://www.kaggle.com/datasets/yelp-dataset/yelp-dataset}}, and MovieLens-25M\footnote{\url{https://grouplens.org/datasets/movielens/25m/}}. \textbf{MovieLens-1M} (ML-1M for short) contains 1 million ratings provided by 6000 users for 4000 movies. \textbf{Yelp} provides real-world data related to businesses, including about 8 million reviews, 209, 393 businesses, and 1, 968, 703 users. \textbf{MovieLens-25M} (ML-25M for short) has 25 million ratings applied to 62,000 movies by 162,000 users. 
Following ~\cite{zhou2018din,xi2023towards}, we convert the ratings into binary labels by labeling ratings of 4 and 5 as positive and the rest as negative and split data into training and test sets in a 9:1 ratio based on user IDs. As described in Section~\ref{sec:llm_size}, our pipeline begins by leveraging a randomly sampled portion of the training set to fine-tune LLMs, improving their representation and compression capabilities. Then, the fine-tuned LLMs are employed to encode user behaviors and item titles to obtain representations. These representations are subsequently used for training the downstream recommendation tasks on the full training set following~\cite{xi2023towards}.

\subsubsection{Baselines and Downstream Models}
In terms of baselines, we first select compression methods that do not fine-tune LLMs, such as \textbf{PCA}~\cite{mackiewicz1993pca}, Autoencoder (\textbf{AE})~\cite{wang2016ae}, and Variational Autoencoder (\textbf{VAE})~\cite{pu2016vae}. Then, we choose several approaches fine-tuning LLMs, including two projection-based compression methods, \textbf{LARR}~\cite{wan2024larr} and \textbf{BAHE}~\cite{geng2024breaking}, and two nested-based approaches, \textbf{ESE}~\cite{wang20242d} and \textbf{MRL}~\cite{kusupati2022matryoshka}. 
LARR~\cite{wan2024larr} applies an MLP to compress representations from LLM and then performs contrastive learning. BAHE~\cite{geng2024breaking} utilizes a linear layer for compression and incorporates a CTR head and cross-entropy loss. ESE~\cite{wang20242d} leverages AoE loss~\cite{li2024aoe} and PCA-like constraints to guide the model in learning compact and efficient representations. MRL~\cite{kusupati2022matryoshka} employs nesting to concentrate critical information in the early part of the representation. The above models, similar to our MARC, are based on Llama3-8B~\cite{llama} and are trained on the same click-through data. Even for contrastive learning-based methods, \eg, LARR, we treat clicked user-item pairs as positive samples, injecting the same supervised signal. Additionally, we include smaller models such as TinyBERT~\cite{jiao2019tinybert}, BERT~\cite{kenton2019bert}, Qwen2-1.5B~\cite{hui2024qwen2}, Phi-2 (2.7B)~\cite{javaheripi2023phi}, and Qwen2-7B~\cite{hui2024qwen2} and utilize their original representations without compression.

For downstream recommendation tasks, we primarily focus on CTR prediction in the ranking stage. The experiments involve several widely adopted CTR models, including \textbf{DCNv2}~\cite{wang2021dcnv2}, \textbf{DCNv1}~\cite{wang2017dcnv1}, \textbf{DIN}~\cite{zhou2018din}, \textbf{DeepFM}~\cite{guo2017deepfm}, and \textbf{AutoInt}~\cite{song2019autoint}. Additionally, to demonstrate MARC's compatibility, we also explore its application in other stages of recommendation, such as re-ranking and retrieval. For the re-ranking stage, we select the state-of-the-art model \textbf{PRM}~\cite{pei2019prm}, while for the retrieval stage, we employ the widely-used \textbf{DSSM}~\cite{huang2013dssm}. 

\subsubsection{Metrics}
We establish the following evaluation metrics based on previous works~\cite{wang2021dcnv2,zhou2018din,pei2019prm,ndcg,yue2007support}.
For CTR prediction tasks, we adopt widely used \textit{AUC} and \textit{Logloss} as metrics. It is worth noting that in CTR prediction, an improvement of 0.001 in AUC is generally considered significant and impactful~\cite{wang2017dcnv1,wang2021dcnv2,zhou2018din}.
For re-ranking tasks, we use \textit{NDCG@K} and \textit{MAP@K} as metrics, where $K=1, 5, 10$. 
For retrieval tasks, we utilize \textit{NDCG@K}, \textit{HitRate@K}, and \textit{MRR} (Mean Reciprocal Rank) as metrics, where $K=10,20,50$.

\begin{table*}[t]

    \vspace{-0pt}
    \caption{The overall performance on the re-ranking task. The symbol * indicates statistically significant improvement over the best compression baselines MRL (t-test with $p < 0.05$).}
    \vspace{-0pt}
    \centering
    \scalebox{1}{
    \setlength{\tabcolsep}{2mm}
{\begin{tabular}{c|cccc|cccc}
\toprule
Models & MAP@1 & MAP@3 & @MAP5 & MAP@7 & NDCG@1 & NDCG@3 & NDCG@5 & NDCG@7 \\
 \midrule
base (PRM) & 0.6535 & 0.7368 & 0.7259 & 0.7212 & 0.6535 & 0.6938 & 0.7652 & 0.7795 \\
Llama3-8B & 0.6789 & 0.7541 & 0.7422 & 0.7388 & 0.6789 & 0.7145 & 0.7814 & 0.7926 \\
MRL & 0.6750 & 0.7505 & 0.7395 & 0.7357 & 0.6750 & 0.7108 & 0.7791 & 0.7906 \\
MARC-C & \textbf{0.6806}* & \textbf{0.7559}* & \textbf{0.7445}* & \textbf{0.7406}* & \textbf{0.6806}* & \textbf{0.7185}* & \textbf{0.7835}* & \textbf{0.7946}* \\
\midrule
MARC-O & 0.6865* & 0.7592* & 0.7475* & 0.7440* & 0.6865* & 0.7222* & 0.7863* & 0.7966*
\\
 \bottomrule
\end{tabular}}
}
\label{tab:overall_rerank}
\vspace{-5pt}
\end{table*}

\subsubsection{Reproducibility}
We randomly sample 200,000 data points from the training set to train the LLM-based representation model. Then, the user and item representations obtained from this LLM are used to train the downstream recommendation task on the full training data. Unless otherwise specified, MARC and compression baselines are based on Llama3-8B, and the downstream task is CTR prediction with \textbf{base} model DCNv2~\cite{wang2021dcnv2}. MARC and compression baselines all produce 128-dimensional compressed representations. \textbf{MARC-O} denotes the original 4096-dimensional representations and \textbf{MARC-C} is the compressed ones. The compression network is implemented as an MLP with hidden layers of [256, 128], while the user-item matching network is an MLP with hidden layers of [128, 1]. The LLM module employs LoRA~\cite{hu2021lora} for fine-tuning and $\alpha$ in Eq.~\eqref{eq:loss} is set to 0.01 to balance the magnitude of losses and we found the model performance is generally robust around this order of magnitude. We follow~\cite{xi2023towards}, leveraging MoE to integrate representations into the downstream model. Each expert in the MoE is an MLP with hidden layer sizes of [128, 32], and the number of experts varies with different downstream tasks and models, typically ranging from 2 to 5. The embedding size for the downstream CTR model is fixed at 32, and the output layer MLP size is [200, 80]. Other parameters, such as batch size and learning rate, are determined with a grid search.  Regarding computational overhead, the training time of MARC is comparable to current fine-tuning-based compression methods as it only introduces lightweight modules.

\subsection{Overall Performance (RQ1)}
\subsubsection{Performance on CTR prediction task.}
We apply the representations from MARC and various baselines to the downstream CTR prediction task on three public datasets, with results shown in Table~\ref{tab:overall}. ``Base'' refers to DCNv2 without representation enhancement. The key findings are as follows:
\textbf{(i)} Our MARC consistently and significantly outperforms other compression models. For example, on ML-1M and Yelp, MARC improves AUC by 0.89\% and 0.74\%, respectively, compared to the best compression baseline MRL. This improvement is due to MARC’s ability, as shown in Section~\ref{sec:mra}, to alleviate mid-layer representation advantage and generate better representations. \textbf{(ii)} MARC produces compressed representations with high information density. MARC-C incurs minimal accuracy loss compared to MARC-O and even outperforms 4096-dimensional representations from frozen Llama3-8B without fine-tuning. For example, on ML-1M, the 128-dimensional MARC-C improves by 0.63\% over representations from frozen Llama3-8B, with only a 0.11\% accuracy loss compared to MARC-O. This suggests that our modular adjustment and task decoupling help the model better retain key information during compression. \textbf{(iii)} Fine-tuning LLMs for compression generally yields better results than non-fine-tuned methods. Fine-tuned approaches, \eg, LARR, MRL, and MARC, outperform non-fine-tuned methods like PCA, AE, and VAE, likely because non-fine-tuned methods struggle to capture information useful for recommendations.

\subsubsection{Performance on re-ranking \& retrieval  tasks}\label{sec:rerank_retrieval}
To investigate whether MARC’s representations are suitable for various recommendation tasks, we apply them to DSSM in the retrieval stage and PRM in the re-ranking stage on the ML-1M dataset. We compare MARC’s original (\textbf{MARC-O}) and compressed (\textbf{MARC-C}) representations against the base model without representation enhancement (\textbf{base}), the 4096-dimensional representations from \textbf{Llama3-8B} without fine-tuning, and the strongest compression baseline, \textbf{MRL}. Both MARC and MRL are fine-tuned on Llama3-8B and generate 128-dimensional compressed representations. The results presented in Table~\ref{tab:overall_rerank} and Table~\ref{tab:overall_retrieval} indicate that representation enhancement significantly improves the base model's performance in both the re-ranking and retrieval stages, aligning with conclusions drawn from CTR prediction tasks. Notably, MARC outperforms both the strongest compression baseline, MRL, and the 4096-dimensional representations from frozen Llama3-8B. For instance, MARC-C improves HitRate@10 in the retrieval task by 12.67\% over MRL and by 8.76\% over frozen Llama3-8B. 
We attribute this generalization to our modular design: by offloading the task adaptation to the Matching Network, the compressed representations are less likely to overfit to the specific objective and instead retain the rich, generalized information beneficial for diverse downstream tasks.

\begin{table}[t]

    \caption{The overall performance on retrieval task. The symbol * indicates statistically significant improvement over the best compression baselines MRL (t-test with $p < 0.05$).}
    \vspace{-5pt}
    \centering
    \scalebox{0.83}{
    \setlength{\tabcolsep}{1mm}
{\begin{tabular}{c|ccc|ccc|c}
\toprule
\multirow{2}{*}{Models} & \multicolumn{3}{c|}{NDCG} & \multicolumn{3}{c|}{HitRate} & MRR \\
\cmidrule{2-8}
 & @10 & @20 & @50 & @10 & @20 & @50 & / \\
 \midrule
base (DSSM) & 0.1570 & 0.1838 & 0.2132 & 0.2867 & 0.3924 & 0.5403 & 0.1324 \\
Llama3-8B & 0.1938 & 0.2195 & 0.2490 & 0.3291 & 0.4308 & 0.5786 & 0.1669 \\
MRL & 0.1842 & 0.2115 & 0.2402 & 0.3177 & 0.4255 & 0.5698 & 0.1582 \\
MARC-C & \textbf{0.2089}* & \textbf{0.2342}* & \textbf{0.2631}* & \textbf{0.3580}* & \textbf{0.4577}* & \textbf{0.6026}* & \textbf{0.1770}* \\
\midrule
MARC-O & 0.2116* & 0.2386* & 0.2666* & 0.3591* & 0.4658* & 0.6066* & 0.1806*
\\
 \bottomrule
\end{tabular}}
}
\label{tab:overall_retrieval}
\vspace{-5pt}
\end{table}

\subsubsection{Online A/B Test (RQ2).}
We applied MARC's representation to the downstream CVR model (conversion rate) in a commercial advertising scenario with tens of millions of users and ads.  The massive scale of users and items makes using raw LLM representations computationally prohibitive for storage and training, hindering real-world deployment. Our model, however, successfully compresses these representations into 128 dimensions with high quality. First, we train our MARC model offline on an industrial dataset to generate compact representations which are updated periodically. These representations are then integrated as features into a downstream CVR model specifically tailored for this scenario.

For the online A/B test, we randomly allocated 10\% of users each to the experimental and control groups. While both groups used the same backbone CVR model, the experimental group integrated MARC representations. In contrast, the control group utilized the current online production champion. It incorporates a highly optimized compression scheme tailored to our industrial-scale environment to reduce the raw LLM representations to 128 dimensions. 
The original, high-dimensional representations are not used online due to their large deployment costs. In a 7-day online A/B test, MARC achieved a \textbf{2.82\% increase in eCPM} against a strong production baseline, with similar inference latency.

\subsection{In-depth Analysis}
\subsubsection{Ablation Study (RQ3)}
\begin{figure}[h]
    \centering
    \vspace{-0pt}
    \includegraphics[width=0.49\textwidth]{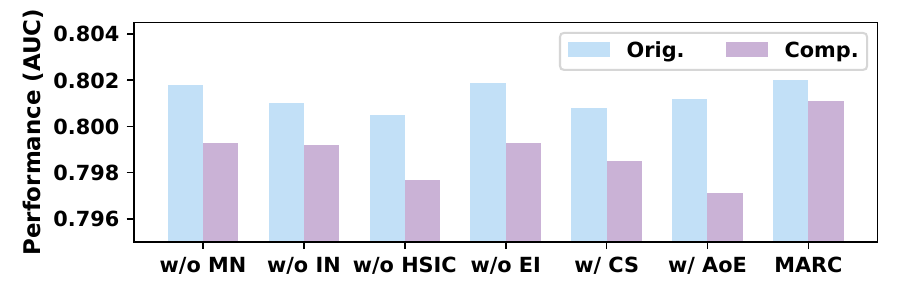}
    \vspace{-10pt}
    \caption{Ablation study on ML-1M dataset.}
    \vspace{-3pt}
    \label{fig:ablation}
\end{figure}

We design several variants and compare their performance in enhancing DCNv2 on the ML-1M dataset. \textbf{w/o IN} removes interaction components from MARC, replacing the user-item matching network with separate MLPs for users and items, and optimized with a cosine similarity loss.
\textbf{w/o MN} removes the matching network and calculates cosine similarity loss directly from compressed representations.
\textbf{w/o EI} removes explicit interactions from Eq.~\eqref{eq:matching_net}.
\textbf{w/o HSIC} removes the HSIC loss.
\textbf{w/ CS} maintains the same network structure but replaces the objective with cosine similarity loss. \textbf{w/ AoE} employs SOTA AoE loss~\cite{li2024aoe} (consisting of angle, cosine similarity, and contrastive objectives) and replaces the matching network with two separate MLPs for user and item because AoE loss is incompatible with it. 
Figure~\ref{fig:ablation} shows the results, where ``\textbf{Orig.}'' denotes original representations and ``\textbf{Comp.}'' refers to compressed ones. 

The results indicate that removing any component leads to a decline in performance. The absence of HSIC loss (w/o HSIC) leads to the largest performance drop for both original and compressed representations, emphasizing its importance for information constraints. Removing explicit interactions (w/o EI) or the matching network (w/o MN) has a limited effect on the original representations but significantly impacts the compressed ones, indicating their role in maintaining information during compression.  However, it is crucial to highlight that even without explicit interactions (w/o EI) or the matching network (w/o MN), MARC still outperforms the strongest baseline MRL. This confirms that while explicit interactions contribute to performance, the core gain stems from our modular decoupling framework. Switching losses (w/ AoE and w/ CS) also degrade representation quality, likely due to their reduced alignment with downstream tasks. AoE performs poorly in compression, likely because it cannot utilize matching network, failing to preserve interaction information crucial for downstream tasks. 

\begin{figure}[htbp]
    \centering
    \vspace{-0pt}
    \includegraphics[width=0.43\textwidth]{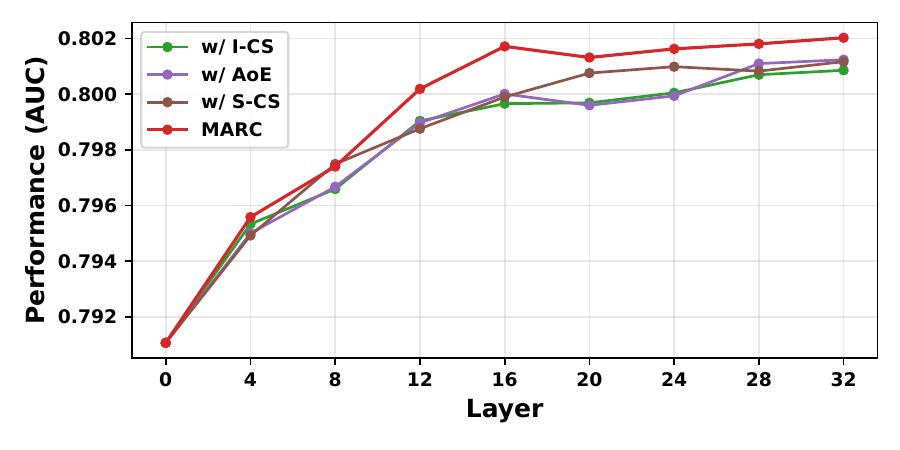}
    \vspace{-5pt}
    \caption{MARC alleviates MRA under various losses.}
    \vspace{-3pt}
    \label{fig:mra_solve}
\end{figure}

Additionally, we evaluate MARC's performance across different loss functions and network architectures to identify what alleviates MRA. We designed three variants: \textbf{w/ I-CS}, which retains the original MARC structure but replaces cross-entropy loss with cosine similarity loss, same as w/ CS; \textbf{w/ AoE}, which employs the state-of-the-art AoE loss. Additionally, we devise \textbf{w/ S-CS}, which also adopts cosine similarity loss, but replaces the interactive matching network with the non-interactive network used in w/ AoE.
The results of these variants' representations on ML-1M are shown in Figure~\ref{fig:mra_solve}. Neither switching loss functions nor removing the interactive component of the matching network yields a mid-layer representation advantage (MRA), yet both still outperform baselines. In all cases, good representations come from the final layer, suggesting that \textbf{the primary factor alleviating MRA in MARC is the modular structure}, rather than the loss function or interactive network. The loss function and the interaction in the matching network primarily enhance representation quality.

\subsubsection{Compatibility Analysis (RQ4)}
\begin{figure}[h]
    \centering
    \vspace{-3pt}
    \includegraphics[width=0.45\textwidth]{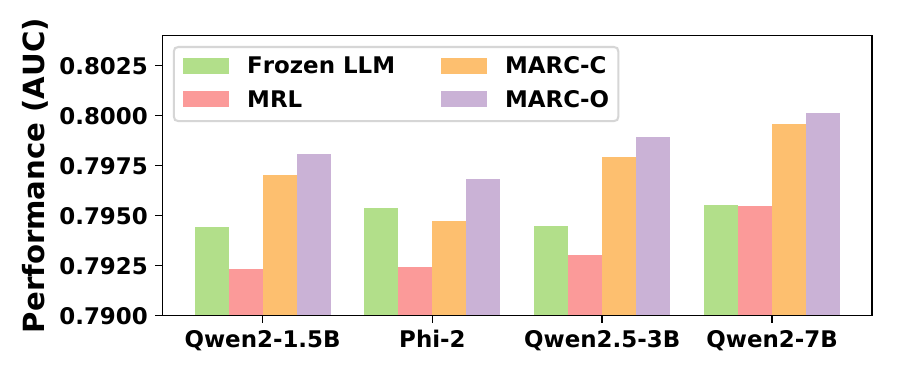}
    \vspace{-3pt}
    \caption{Performance on different backbone LLMs. Frozen LLM denotes backbone LLM without fine-tuning.}
    \vspace{-3pt}
    \label{fig:diff_llms}
\end{figure}

Previous experiments have explored the compatibility of MARC across various datasets and downstream recommendation tasks. This section delves into the compatibility of MARC with different CTR models and backbone LLMs. We select several commonly used CTR models, including \textbf{DCNv1}~\cite{wang2017dcnv1}, \textbf{DeepFM}~\cite{guo2017deepfm}, \textbf{FiGNN}~\cite{li2019fignn}, \textbf{FiBiNet}~\cite{huang2019fibinet}, \textbf{AutoInt}~\cite{song2019autoint}, \textbf{xDeepFM}~\cite{lian2018xdeepfm}, and \textbf{DIN}~\cite{zhou2018din}, and the results are presented in Table~\ref{tab:ctr_models_1}. Regarding backbone LLMs, we select a range of models with different sizes, including \textbf{Qwen2-1.5B}, \textbf{Phi-2} (2.7B), \textbf{Qwen2.5-3B}, and \textbf{Qwen2-7B}, with their results based on DCNv2 shown in Figure~\ref{fig:diff_llms}. As in Section~\ref{sec:rerank_retrieval}, we compare MARC's original and compressed representations, MARC-O and MARC-C, with the high-dimensional representations from backbone LLMs without fine-tuning (\textbf{Frozen LLM}) and 128-dimensional representations from the strongest compression baseline, \textbf{MRL}.


 

\begin{table*}[h]

    \vspace{-0pt}
    \caption{The overall performance on different CTR models.}
    \vspace{-5pt}
    \centering
    \scalebox{0.9}{
    \setlength{\tabcolsep}{1.5mm}
{\begin{tabular}{c|cccccccccccccc}
\toprule
\multirow{2}{*}{\textbf{Models}} & \multicolumn{2}{c}{xDeepFM} & \multicolumn{2}{c}{FiBiNet} & \multicolumn{2}{c}{DeepFM} & \multicolumn{2}{c}{FiGNN} & \multicolumn{2}{c}{DCNv1} & \multicolumn{2}{c}{DIN} & \multicolumn{2}{c}{AutoInt} \\
\cmidrule{2-15}
 & AUC              & Logloss           & AUC              & Logloss           & AUC         & Logloss      & AUC         & Logloss     & AUC         & Logloss     & AUC        & Logloss    & AUC          & Logloss      \\
 \midrule
base                             & 0.7830           & 0.5514            & 0.7824           & 0.5518            & 0.7827      & 0.5514       & 0.7838      & 0.5502      & 0.7837      & 0.5515      & 0.7887     & 0.5450     & 0.7827       & 0.5515       \\
Llama3-8B                        & 0.7962           & 0.5378            & 0.7963           & 0.5398            & 0.7956      & 0.5431       & 0.7966      & 0.5398      & 0.7961      & 0.5390      & 0.7983     & 0.5367     & 0.7965       & 0.5434       \\
MRL                              & 0.7958           & 0.5384            & 0.7956           & 0.5384            & 0.7945      & 0.5397       & 0.7959      & 0.5383      & 0.7956      & 0.5400      & 0.7972     & 0.5368     & 0.7958       & 0.5392       \\
MARC-C                           & \textbf{0.8009}           & \textbf{0.5336}            & \textbf{0.8013}           & \textbf{0.5315}            & \textbf{0.8001}      & \textbf{0.5332}       & \textbf{0.8012}      & \textbf{0.5323}      & \textbf{0.8016}      & \textbf{0.5329}      & \textbf{0.8019}     & \textbf{0.5323}     & \textbf{0.8006}       & \textbf{0.5351}       \\
\midrule
MARC-O                           & 0.8024           & 0.5348            & 0.8023           & 0.5310            & 0.8017      & 0.5320       & 0.8020      & 0.5309      & 0.8021      & 0.5307      & 0.8021     & 0.5307     & 0.8033       & 0.5303     \\
\bottomrule

\end{tabular}}
}
\label{tab:ctr_models_1}
\vspace{-5pt}
\end{table*}

From Table~\ref{tab:ctr_models_1}, it is evident that MARC demonstrates performance similar to its results in DCNv2 across different CTR models. MARC-C significantly outperforms MRL and the frozen Llama3-8B, with a minimal gap of less than 0.1\% when compared to the upper-bound MARC-O. Additionally, Figure~\ref{fig:diff_llms} demonstrates that MARC also exhibits superior performance with various backbone LLMs, although the effectiveness of the representation may vary depending on the size of the LLMs. Furthermore, we verify that MARC can mitigate mid-layer representation advantage across different downstream models and LLMs, which is detailed in Section~\ref{sec:mra}. This confirms that our modular compression framework is versatile and compatible with a wide range of downstream models and backbones, alleviates mid-layer representation advantage, and yields lightweight yet efficient representations.

\subsubsection{Impact of Compressed Representation Dimension (RQ5)} 
\begin{figure}[h]
    \centering
    \vspace{-3pt}
    \includegraphics[width=0.45\textwidth]{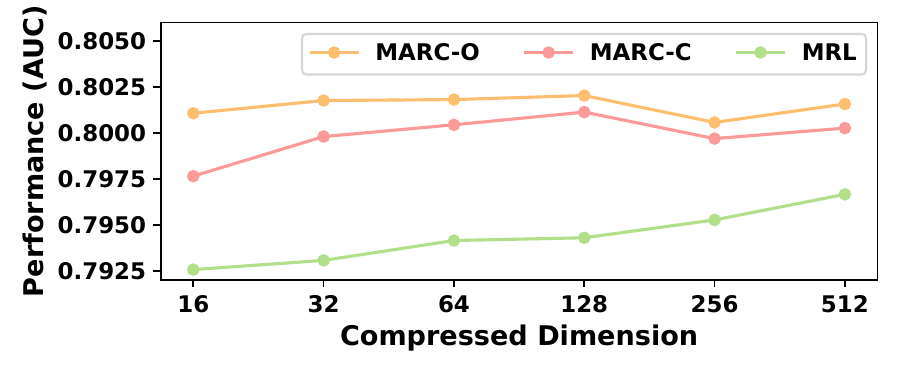}
    \vspace{-3pt}
    \caption{Performance under different dimensions.}
    \vspace{-3pt}
    \label{fig:diff_dims}
\end{figure}

Lastly, we investigate the impact of the compressed representation dimensions on the representation quality. We vary the dimensionalities of compressed representations from MARC and the strongest baseline MRL, including 16, 32, 64, 128, 256, and 512, and apply them to the downstream DCNv2 model. The results are presented in Figure~\ref{fig:diff_dims}, where MARC-C and MARC-O represent the compressed and original representations from MARC, respectively. From the figure, it is evident that MARC significantly outperforms the baseline MRL across all dimensions, demonstrating that our model can generate more efficient and lightweight representations. When the compressed dimension is small, the original representation performs better. As the dimension increases, compressed representations demonstrate improved performance, likely because larger dimensions retain more information. However, after the dimension of 128, the performance of MARC-O begins to decline, and since the original representations serve as the upper bound for compressed representations, this also leads to a decrease in the performance of MARC-C. This decline may be attributed to the increasing difficulty of model training as the compressed dimension grows larger.

\section{Related Work}
\subsection{Text Embedding and Compression}
With the rise of LLMs, many efforts in NLP have focused on leveraging LLMs for sentence embedding learning. Most of these approaches utilize unsupervised contrastive learning, which contrasts in-batch positive and negative samples to improve representation quality~\cite{behnamghader2024llm2vec,xu2024bmretriever,muennighoff2024generative,lee2024gecko,li2024bellm,xi2025infodeepseek}. Some supervised approaches have also been proposed to align embeddings with human perception, typically by maximizing the cosine similarity between positive pairs~\cite{li2024aoe,jiang2023scaling,wang20242d}. However, these methods often take the pooling of the last hidden states of LLMs as the final representation, resulting in a high-dimensional representation (\eg, 4096). Such representations pose challenges for storage and downstream tasks, especially under resource constraints~\cite{wang20242d}. To address this, nested-based representation compression has been introduced in NLP, including Matryoshka Representation Learning (MRL)~\cite{kusupati2022matryoshka,lee2024gecko,ma2024fine} and PCA-like approaches~\cite{wang20242d}. MRL learns nested lower-dimensional vectors within the same high-dimensional space to preserve information at multiple granularities, while PCA-like approaches apply PCA during training to concentrate important information in the leading dimensions of the representation. There are also some projection-based approaches that employ simple projection layers to compress the final representations~\cite{wan2024larr,geng2024breaking}. Note that we mainly focus on representation compression that directly maps high-dimensional LLM vectors to low-dimensional spaces. We do not include knowledge distillation approaches, which belong to model compression.

The above representation compression methods often suffer from the mid-layer representation advantage, where intermediate representations perform better than final ones, leading to suboptimal compressed embeddings, as demonstrated in Section~\ref{sec:mra}. Our work focuses on addressing this issue to enhance representation quality.

\subsection{LLM-based Recommendation}
The emergence of LLMs has brought significant transformations to RSs~\cite{liu2024mamba4rec,li2023large,zhu2023large,chen2023large,lin2024clickprompt,wu2023survey,liu2023pre,yu2023self,xi2024memocrs,xi2024play}, with two main approaches: LLMs as core recommenders and as parts of traditional ones. The former uses LLMs as recommenders, which leads to superior performance, but faces deployment challenges due to large inference latency~\cite{bao2023tallrec,zheng2024adapting,zhu2024collaborative,zheng2024harnessing,dong2024unsupervised,tan2024idgenrec,xi2025bursting}. The latter integrates LLMs into traditional RSs, using LLM-derived representations for downstream RSs. Here, some methods first generate recommendation-related knowledge with LLMs and then encode this knowledge into representations~\cite{xi2023towards,lyu2023llm,ren2024representation,liu2024once,tian2024reland,xi2024decoding,xi2024efficient}, while others directly encode recommendation-related text with LLMs~\cite{zhang2024notellm,wan2024larr,jia2024knowledge,geng2024breaking,wang2024can}. Regardless of the specific method, leveraging LLMs' representations is a crucial step.

However, current deployable solutions still face notable challenges. Approaches that utilize smaller models~\cite{xi2023towards,luo2024kellmrec,liu2024once,tian2024reland,wang2024can}, \eg, BERT, often suffer from significant performance degradation. On the other hand, directly employing LLMs for representation leads to substantial resource consumption~\cite{zhang2024notellm,hu2024enhancing}. Therefore, this work primarily focuses on representation compression, aiming to obtain lightweight and effective RSs representations from LLMs.

\section{Conclusion}

This work identifies a trade-off between efficiency and effectiveness in LLM-based representations for recommendations, and focuses on representation compression. We reveal the phenomenon of mid-layer representation advantage during compression and explain it using modularity theory, where LLMs develop internal functional modularity during training. To address this, we propose MARC, which includes Modular Adjustment and Modular Task Decoupling. Extensive experiments validate that MARC effectively addresses mid-layer advantage and produces efficient representations for recommendation tasks. It also improves eCPM by 2.82\% during an online A/B test in a large commercial search advertising scenario.

\begin{acks}
    The Shanghai Jiao Tong University team is partially supported by Shanghai Municipal Science and Technology Major Project (2021SHZDZX0102), National Natural Science Foundation of China (624B2096, 72595872, 62322603), and National Key RD Program of China (2022ZD0114804), Changan Automobile \& Chongqing Natural Science Foundation Joint Fund for Innovation and Development (CSTB2023NSCQ-LZX0136). The work is also sponsored by Huawei Innovation Research Program. We thank MindSpore~\cite{mindspore} for its partial support. The author Yunjia Xi is also supported by Wu Wen Jun Honorary Doctoral Scholarship.
\end{acks}

\bibliographystyle{ACM-Reference-Format}
\bibliography{sample-base}

\balance

\end{document}